\documentstyle[preprint,aps]{revtex}
\begin{document}
\draft

\title{Self-Consistent Model of Polymerization-Induced \\Phase Separation}

\author{Valeriy V. Ginzburg and Noel A. Clark}
\address{ Department of Physics, University of Colorado,\\
 Boulder CO 80309-0390 USA}
\date{}
\maketitle

\begin{abstract}We propose an exactly solvable self-consistent kinetic model of
polymerization-induced phase separation (PIPS) via spinodal decomposition.
Using modified Cahn-Hilliard 
and Glotzer-Coniglio theories for early and late stages of spinodal decomposition, 
we find scaling regimes and compare the obtained results with existing experimental
and theoretical information on PIPS kinetics, finding a qualitative agreement.
\end{abstract}


\section{Introduction}

In the last several years, polymer-dispersed liquid crystals (PDLCs) have generated
significant interest among scientists because of their utility for display
applications (see, e.g., \cite{Doane1} - \cite{Drzaic}). Among the most important problems
in PDLC preparation is to control the phase separation between the polymer and the liquid 
crystal. It is known that such phase separation can be achieved either by performing
polymerization in the monomer - liquid crystal (LC) mixture at high temperature (where
polymer and LC are miscible) and then quenching to low temperatures and thus inducing
phase separation [temperature induced phase separation (TIPS)], or by performing polymerization
at such temperature at which monomer and LC are miscible, yet polymer and LC are immiscible
[polymerization induced phase separation (PIPS)]. In the former case, the phase separation is
controlled by the quench depth, in the latter case, it is controlled by the polymerization
kinetics.

Recently, several studies of PIPS kinetics, experimental~(\cite{Palffy1} - \cite{Kyu1}),
theoretical~\cite{Taylor}, and numerical~(\cite{Teixeira} - \cite{Lansac}), have been performed.
These results showed that PIPS is reasonably well described by theories of ordinary phase
separation induced by a thermal quench, so that at the early stage of phase separation (PS),
classical linear theory of Cahn and Hilliard~(\cite{Cahn1} -\cite{Cahn3}) is applicable for
the initial growth of phase regions; at later stages, scaling models of Binder and 
Stauffer~\cite{Binder}, Langer, Bar-on and Miller~\cite{Langer} and 
Siggia~\cite{Siggia} are used to describe the behavior of scattering function or domain 
size. At the same time, no model exists yet that describes the analytical dependence of PS
kinetics on polymerization rate. The
purpose of this work is to provide at least a qualitative approach to the problem.

The paper is organized as follows: in section~\ref{sec:td} we introduce thermodynamic and
kinetic variables used in the problem, discuss the phase diagram and make approximations
about conditions of polymerization. In section~\ref{sec:kineq} we derive hydrodynamic
equation of PS and discuss the self-consistent approximation. Then, in section~
\ref{sec:regimes}, we analyze approximate scaling solutions of these kinetic equations and
compare them with existing experimental and theoretical information. Finally, in section
~\ref{sec:concl}, we discuss the validity of the approximations used and analyze possible
practical use of the model and future changes and improvements.

\section{Phase Diagram and Thermodynamic Variables}
\label{sec:td}

We will consider a system consisting of monomer, polymer and LC in isotropic state. 
Following Teixeira and Mulder\cite{Teixeira}, we will neglect polydispersity of polymer
lengths and disregard any difference between gelation and polymerization. In this case,
the system under consideration can be described as a ternary mixture with $\phi$ being the
volume fraction of LC, M being the volume fraction of the monomer, and P being the volume
fraction of polymer. We assume an incompressibility condition that reads:

\begin{equation}
	\phi + M + P = 1.
\label{eq:incompr}
\end{equation}

Because of condition~(\ref{eq:incompr}), only two of the three concentrations are 
independent, and it is more convenient to work with variables $\phi$ and 
$Y = P/(P+M)$ - the degree of conversion. We assume that polymer and monomer are
always miscible, and that the monomer - LC interaction and the polymer segment - LC interaction
are the same. In this case, the Flory-Huggins free energy of the mixture can be written as:

\begin{equation}
	F = \int dV [ \phi \ln \phi + (1-Y)(1-\phi) \ln (1-\phi) + \chi \phi (1-\phi) +
	\frac{\kappa}{2}(\nabla \phi)^{2} ],
\label{eq:fh}
\end{equation}

where $\chi$ is Flory-Huggins parameter, related to temperature (it is usually assumed that
$\chi = A + B/T$, where $B$ is a positive constant, $A$ is an arbitrary constant).

The schematic phase diagram for the system~(\ref{eq:fh}) is shown in Figure 1, with different
curves showing spinodal lines for different values of conversion parameter $Y$. It is similar
to spinodals of polymer-LC system derived in~\cite{Kyu2}. It can
be seen that with Y increasing, the critical temperature and critical concentration move
toward high temperatures and low concentrations of polymer, respectively. For each
value of the Flory-Huggins parameter $\chi$ between $\chi_{c1}$ and $\chi_{c2}$, 
there exists a critical concentration $\phi_{c}(\chi)$ and a critical degree of 
conversion $Y_{c}(\chi)$, so that the point
$(\chi, \phi_{c}(\chi), Y_{c}(\chi))$ is a critical point. 

We will assume that polymer - LC phase separation occurs only via the spinodal 
decomposition (SD) mechanism and neglect nucleation in the metastable phase. Further, we
will assume that the temperature at which the process occurs, is not very far below the
critical polymer-LC temperature. In that case, as it will be shown later in this section,
expression for the free energy used in deriving kinetic equations, can be simplified
dramatically. 

For a given $\chi$, the critical composition $\phi_{c}(\chi)$ and the critical conversion 
$Y_{c}(\chi)$ are:

\begin{equation}
	Y_{c}(\chi) = \sqrt{8 \chi} - 2 \chi,
\label{eq:ycr}
\end{equation}

\begin{equation}
	\phi_{c} = \sqrt{ \frac{1}{2 \chi} }.
\label{eq:phicr}
\end{equation}

The isothermal (fixed $\chi$) spinodal curve, such as shown in Figure 2, can be described by the following equation:
\begin{equation}
	\phi - \phi_{c} \approx \frac{Y - Y_{c}}{4 \chi} \pm \sqrt{ \frac{ (Y-Y_{c})(Y+Y_{c}+4 \chi) }{16 \chi^{2}}} \approx \sqrt{ \frac{Y-Y_{c}}{4 \chi}},
\label{eq:spinexact}
\end{equation}
the second equality valid in the limit $Y, Y_{c} << 4 \chi$.
We then construct the Ginzburg-Landau free energy density 
corresponding to this modified spinodal (i.e., all minima of this free energy expression lie on the spinodal):

\begin{equation}
	f_{GL} = - \frac{Y-Y_{c}}{2} X^{2} + \chi X^{4},	
\label{eq:landau}
\end{equation}

\begin{equation}
	\beta F = \int dV [ f_{GL} + \frac{\kappa}{2}(\nabla \phi)^{2} ],
\label{eq:lgfreen}
\end{equation}

where $X = \phi - \phi_{c}$.

In the next section, we will derive kinetic equations and analyze their solutions in the self-consistent
approximation.

\section{Kinetic Equations}
\label{sec:kineq}

In order to fully describe PIPS kinetics, it is necessary to determine time dependence of both degree of
conversion $Y$ and phase separation $X$. The former variable is a nonconserved, the latter one is a conserved
order parameter. Assuming that polymerization occurs as a simple bimolecular reaction with a rate ${\cal K}$, we
can write equation for $Y$ as follows:

\begin{equation}
	\frac{\partial Y}{\partial t} = {\cal K} (1 - Y)^{2} (1 - \phi_{c} - X)^{2}.
\label{eq:reactionexact}
\end{equation}

For the phase separation, we employ the classical time-dependent Ginzburg-Landau (TDGL) equation:

\begin{equation}
	\frac{\partial X}{\partial t} = \Gamma \nabla^{2} [ \frac{\delta f_{GL}}{\delta X} - \kappa \nabla^{2} X ].
\label{eq:tdgl}
\end{equation}

As written, equations~(\ref{eq:reactionexact}) - (\ref{eq:tdgl}) constitute a very complex system of coupled nonlinear
partial differential equations, susceptible only to limited numerical treatment. We make several approximations in
order to somewhat simplify them and make them suitable for analytical study. First, we average spatially the equation~
(\ref{eq:reactionexact}) to obtain a spatially independent $Y(t) = <Y({\bf r},t)>$:

\begin{equation}
	\frac{d Y}{d t} = {\cal K} (1 - Y)^{2} [(1 - \phi_{c})^{2} + <X^{2}>].
\label{eq:reactionaver}
\end{equation}
	
The system of coupled equations~(\ref{eq:tdgl} - \ref{eq:reactionaver}) somewhat resembles the kinetic equations that
we derived for the annihilation-diffusion systems:

\begin{equation}
	\frac{d \rho}{d t}  = - {\cal K} (\rho^{2} - <f^{2}>)), 
\label{eq:rhoevol}
\end{equation}

\begin{equation}
	\frac{\partial f({\bf r},t)}{\partial t} - D \nabla^{2} f({\bf r},t) = 0, 
\label{eq:fevol}
\end{equation}

where the (averaged) particle density $\rho$ is a nonconserved order parameter decaying to 0 during the
annihilation process, and the charge density $f$ is a conserved order parameter whose fluctuations play
important role in the annihilation process~\cite{Ginzburg96}. However, the PIPS kinetics is more 
complicated than the annihilation-diffusion process because equation~(\ref{eq:tdgl}) is nonlinear in $X$.
On the other hand, it seems that in polymerization, fluctuations play a less important role than in annihilation,
and it is possible to decouple the two by neglecting the $<X^{2}>$ term in (\ref{eq:reactionaver}) and essentially
having $Y(t)$ a known function to be plugged in equation (\ref{eq:tdgl}). Therefore, we will not consider 
equation (\ref{eq:reactionaver}) in our subsequent analysis and instead concentrate only on the equation 
~(\ref{eq:tdgl}).

We will also assume critical polymerization, i.e., that the system enters the two-phase region on a phase
diagram through a critical point, undergoing a second-order transition. Mathematically, this means that $<X({\bf r},t)> = 0$. If this is the case,
PIPS occurs in two stages: first, when the concentration of polymer is small, and the mixture is still in the one-
phase region, kinetics is diffusive, fluctuations are generated spontaneously due to random forces, and gradually
shrink due to diffusion; second, when the degree of conversion becomes larger than the critical value $Y_{c}$,
kinetics changes from diffusion to spinodal decomposition, and fluctuations begin to grow rapidly. This (rather
trivial) observation is supported by experimental evidence that PIPS has a relatively large induction time.

In the remainder of this section, we will study in detail the phase separation stage of the process. 
Let us rewrite equation (\ref{eq:tdgl}) in Fourier space and simultaneously transform everything to
dimensionless variables:

\begin{equation}
	\frac{\partial X({\bf k},t)}{\partial t} = - k^{2} [ - \frac{Y - Y_{c}}{1 - Y_{c}} + <X^{2}(t)> + k^{2}]
	X({\bf k},t)
\label{eq:sctdgl}
\end{equation}

where $X \longrightarrow X \sqrt{\frac{4 \chi}{1 - Y_{c}}}$, $k \longrightarrow k \sqrt{\frac{\kappa}{1 - Y_{c}}}$,
$t \longrightarrow t \frac{\Gamma (1 - Y_{c})^{2}}{\kappa}$. We also substituted $X^{3}$ with $<X^{2}>X$, as is usually
done in self-consistent theories (\cite{Coniglio}, \cite{Glotzer}, \cite{Castellano}). Equation (\ref{eq:sctdgl}) can be readily analyzed if one
knows the t-dependence of $Y$. This analysis will be done in the next section.

\section{Solutions of Kinetic Equations - Two Regimes}
\label{sec:regimes}

\subsection{Early Stage of Spinodal Decomposition}

In the early stage of spinodal decomposition, fluctuations are relatively small, and it is possible to linearize the equation (\ref{eq:sctdgl}) by neglecting the $<X^{2}>$ term. Such linearization is used in the original Cahn-Hilliard theory of phase separation after quench into the two-phase region. In PIPS case, however, the system moves into the two-phase region as polymerization proceeds, and so the transition happens at some time $t_{0} > 0$. Here we will not attempt to determine $t_{0}$, but in principle, it is not difficult if polymerization rate is known: at $t=t_{0}$, $Y = Y_{c}$. For the early stage of the process, then, we can write:

\begin{equation}
	\frac{Y-Y_{c}}{1-Y_{c}} = {\cal K} (t - t_{0}).
\label{eq:yoft}
\end{equation}

At later times ($t-t_{0} > {\cal K}^{-1}$), we just assume $\frac{Y-Y_{c}}{1-Y_{c}} = 1$.

The linearized TDGL equation has as its solutions all excitations of the form:

\begin{equation}
	X({\bf k},t) = X({\bf k},t_{0}) \exp [ \nu_{k} (t-t_{0}) ],
\label{eq:linsol}
\end{equation}

with the dispersion relation:

\begin{equation}
	\nu_{k} = 2 k^{2} (\frac{Y-Y_{c}}{2(1-Y_{c})} - k^{2}) = 
	2 k^{2} (\frac{{\cal K} (t-t_{0})}{2} - k^{2}).
\label{eq:dispersion}
\end{equation}

Excitations that grow most rapidly have a wavevector $k_{m}$ given by:
\begin{equation}
	k_{m} = \sqrt{ \frac{{\cal K} (t-t_{0})}{4} }.
\end{equation}

Unlike the thermally induced phase separation, where $k_{m}$ is constant during the early stage, in PIPS it changes with time even in the early stage. The fact that $k_{m}$ increases with time indicates that the predominant mechanism is the formation of new fluctuations rather than the growth of existing ones.

At time $t-t_{0} = {\cal K}^{-1}$, the system have polymerized completely and yet it is still in the early stage of SD. From now on, the regular framework of SD theories is applicable. In the next subsection, we will use the self-consistent approximation with the scaling ansatz to describe the late stage of the SD in PIPS.

\subsection{Late Stage of the Spinodal Decomposition}

In the late stage of SD, we can no longer neglect the $<X^{2}>$ term, and equations of evolution become strongly nonlinear. Even though the self-consistent approximation allows iterative treatment, its complicated character prevents us from writing an exact solution in analytical form. Coniglio and Zannetti~\cite{Coniglio} described the solution of TDGL in this self-consistent approximation to find that $k_{m} \propto (t-t_{0})^{-1/4}$ (we omit logarithmic correction to the power law). Such behavior is called multiscaling and is exact only for the infinite-component order parameter; for the scalar order parameter, one should expect a slightly different exponent. The other scaling theories of PS, based on different mechanisms of domain growth, proposed exponents ranging from 1/3 (Lifshitz-Slyozov \cite{Lifshitz} law, based on a coagulation model) to 0.21 (Binder and Stauffer model \cite{Binder}), so the mean-field value of 0.25 seems to work rather well. We will outline another way of solving the TDGL in self-consistent approximation, and in the process will calculate the scaling function for the structure factor.

Let us start from the scaling ansatz for the structure factor $S({\bf k},t) = <X({\bf k},t) X({\bf -k},t)>$:

\begin{equation}
	S({\bf k},t) = k_{m}^{-d}{\cal F}(k/k_{m}),
\label{eq:ansatz}
\end{equation}

where ${\cal F}$ is a scaling function which has to be determined. Since 

\begin{math}
	<X^{2}(t)> = \int \frac{d^{d} {\bf k}}{(2 \pi)^{d}} S({\bf k},t),
\end{math}

equation (\ref{eq:sctdgl}) can be rewritten in terms of ${\cal F}$ as:

\begin{equation}
	-dk_{m}^{-d-1} \frac{dk_{m}}{dt} {\cal F} - k_{m}^{-d-1}\frac{d {\cal F}}{dx}x \frac{dk_{m}}{dt} = - 2k_{m}^{2-d} x^{2}(-1 + <X^{2}> + k_{m}^{2} x^{2}){\cal F},
\label{eq:kmandf}
\end{equation}

where $x = k/k_{m}$. The scaling function ${\cal F}$ is time independent, and from this requirement we determine the critical exponent $\lambda$ ($k_{m} \propto (t-t_{0})^{- \lambda}$). First, we separate variables in equation~(\ref{eq:kmandf}) to obtain the implicit expression for ${\cal F}$:

\begin{equation}
	\frac{d \ln {\cal F}}{d \ln x} = -d + 2 k_{m}^{3} (\frac{d k_{m}}{d t})^{-1} x^{2} (-1 + <X^{2}>) + 2 k_{m}^{5} (\frac{d k_{m}}{d t})^{-1} x^{4}.
\label{eq:fofx}
\end{equation}

From the requirement that function ${\cal F}$ has a maximum at $x=1$ (stemming simply from the definition of $k_{m}$), we obtain:

\begin{equation}
-d + 2 k_{m}^{3} (\frac{d k_{m}}{d t})^{-1} (-1 + <X^{2}>) + 2 k_{m}^{5} (\frac{d k_{m}}{d t})^{-1} = 0.
\label{eq:odeforkm}
\end{equation}

Moreover, all three terms in equation~(\ref{eq:odeforkm}) should be time-independent, otherwise the scaling ansatz is not applicable. This immediately dictates the critical exponent $\lambda$ equals 1/4, as it is required to be in self-consistent theories of TDGL (see, e.g., Coniglio et al. \cite{Coniglio}). If we attempt to make $k_{m}$ continuous by connecting the early stage asymptotics and the late stage asymptotics at point $t-t_{0} = {\cal K}^{-1}$, we obtain the zero-order approximation for $k_{m}(t)$ , $<X^{2}(t)>$, and ${\cal F}(x)$:

\begin{equation}
	k_{m}^{4} \approx \frac{1}{16 {\cal K} (t - t_{0})};
\label{eq:kmoft}
\end{equation}

\begin{equation}
	<X^{2}> \approx 1 - \frac{d \sqrt{\cal K}}{2 \sqrt{t - t_{0}}};
\label{eq:x2oft}
\end{equation}

\begin{equation}
	\frac{d \ln {\cal F}}{d \ln x} = - \frac{x^{4}}{2 {\cal K}} + [d + \frac{1}{2 {\cal K}}] x^{2} - d,
\label{eq:fofx2}
\end{equation}

with normalization:

\begin{equation}
	\int \frac{d^{d} {\bf x}}{(2 \pi)^{d}} {\cal F}(x) = 1.
\label{eq:normf}
\end{equation}

Equation~(\ref{eq:fofx2}) can be solved to yield the explicit dependence of $\cal F$ on $x$:
\begin{equation}
	{\cal F} = {\cal F}_{0} x^{-d} \exp [- \frac{x^{4}}{8 \cal K} + (d + \frac{1}{2 \cal K}) \frac{x^{2}}{2}],
\label{eq:fofx3}
\end{equation}
where ${\cal F}_{0}$ is simply a normalization constant.

In Figure 3, we compare our scaling function~(\ref{eq:fofx3}) with the one experimentally measured by Palffy-Muhoray et al. \cite{Palffy2}. It can be seen that the proposed scaling function describes the long-wavelength part of the spectrum reasonably well, while in the short-wavelength region, it decays much faster than the experimental scaling function. This behavior is somewhat expected because the mean-field Landau-Ginzburg treatment fares poorly when fluctuations are large and/or domain walls are too steep. An oversimplified treatment of polymerization process may also contribute to this discrepancy. However, agreement in the long-wavelength region, achieved with the help of just one adjustable parameter $\cal K$, is encouraging.

Obviously, solutions (\ref{eq:kmoft}) - (\ref{eq:fofx2}) can be considered only the zero-order approximation even in the self-consistent model. Nonlinear differential equations of evolution should be solved iteratively, and in this case, we expect to find logarithmic corrections to $k_{m}(t)$ and significant changes in the behavior of $<X^{2}>$. Calculated self-consistent solutions thus should be viewed as a qualitative, rather than quantitative, description of the PS process and the influence of polymerization on its kinetics.

The above-described late stage of the SD corresponds to a truly fractal system. It can be viewed as a set of LC-rich  droplets in a polymer-rich matrix; within each droplet there are smaller polymer-rich droplets, and so on. This pattern also has been observed in the PIPS experimental studies \cite{Palffy2}. 

At even larger times, the fractal picture is destroyed. As droplets grow larger, their interaction and overlap, which is not taken into account in the above treatment, becomes more and more important. The droplet size grows not only due to the SD mechanism, but also (and most importantly) due to aggregation and coagulation. Such a mechanism was described by Siggia~\cite{Siggia} in his scaling theory, which yields $\lambda = 1$ at this very late stage of phase separation (such behavior was indeed observed for PIPS in the same experiment \cite{Palffy2}.) However, this stage is not considered in our theory, which was primarily aimed at analyzing the influence of polymerization on the phase separation kinetics. It is rather obvious that at the time when Siggia mechanism becomes dominant, no polymerization takes place, and phase separation occurs in the same way for both PIPS and TIPS.

\section{Conclusions}
\label{sec:concl}

We analyze in a self-consistent approximation the critical polymerization-induced phase separation (PIPS). Contrary to the temperature-induced phase separation (TIPS), in the critical PIPS the phase separation starts only after some induction period, when polymerization advances enough to make liquid crystal strongly incompatible with the polymer. By assuming a stepwise polymerization rate and neglecting the feedback between phase separation and polymerization, we simplified the problem and obtained analytical solution for the early (Cahn-Hilliard) and late (Coniglio-Zannetti) stage of the critical spinodal decomposition (SD). 

Among the other approximations: the isothermal process occurs at relatively high (close to the miscibility of polymer and liquid crystal) temperature - this approximation justifies the use of symmetric Ginzburg-Landau free energy; constant, rather than concentration-dependent, mobility parameter $\Gamma$; gelation is deemed unimportant or non-crosslinked polymer is considered. All these features make model very simple in comparison with real PDLC systems; at the same time, it obviously enables us to obtain a better insight into this complicated process using analytical, rather than purely numerical, method.

Obvious goals for the improvement of the theory would be: expanding the model for the case of non-critical PIPS; allowing for the more realistic description of the polymerization process; accounting for the difference between gelation and polymerization. These studies will be undertaken in the future.

\vspace{0.5in}
{\bf Acknowledgment}. This work was supported by the NSF Grant DMR 92-24168.

\begin{figure}
Figure 1. Spinodals of polymer-monomer-liquid crystal mixtures. Each curve corresponds to a given
degree of conversion from monomer to polymer, from $Y=0$ (no polymer) to $Y=1$ (no monomer left). The
solid horizontal line is a cross-section $\chi = 1.5$.

\vspace{0.5in}

Figure 2. Spinodal points for polymer-monomer-liquid crystal mixtures at fixed temperature ($\chi = 1.5$).

\vspace{0.5in}

Figure 3. The scaling function $F(q/q_{max})$ vs the scaled wavevector $q/q_{max}$. The data (six differentsets of points) are reprinted from Ref.~\cite{Palffy2}; the solid line represents the best fit for these
data using Eq. (\ref{eq:fofx3}) with adjustable parameter ${\cal K} = 0.025$.
\end{figure}


\begin{thebibliography}{99}

\bibitem{Doane1} D. K. Yang, L. C. Chien, and J. W. Doane, Appl. Phys. Lett. {\bf 60}, 3102 (1992).

\bibitem{Doane2}J. W. Doane, N. A. Vaz, B.-G. Wu, and S. Zumer, Appl. Phys. Lett. {\bf 48}, 269 (1986).

\bibitem{Doane3}J. W. Doane, Mat. Res. Soc. Bull. {\bf XVI}, 24 (1991).

\bibitem{Hikmet1}R. A. M. Hikmet, J. Appl. Phys. {\bf 68}, 4406 (1990).

\bibitem{Hikmet2}R. A. M. Hikmet, Liq. Crystals {\bf 9}, 405 (1991).

\bibitem{Guymon}C. A. Guymon, E. N. Hoggan, D. M. Walba, N. A. Clark, and C. N. Bowman,
 Liq. Cryst. {\bf 19}, 719 (1995).

\bibitem{Drzaic}P. S. Drzaic, Liquid Crystal Disperions, World Scientific, 1995, and references
therein.

\bibitem{Palffy1}J. Y. Kim and P. Palffy-Muhoray, Mol. Cryst. Liq. Cryst. {\bf 203}, 93 (1991).

\bibitem{Palffy2}J. Y. Kim, C. H. Cho, P. Palffy-Muhoray, M. Mustafa, and T. Kyu, Phys. Rev. Lett.
 {\bf 71}, 2232 (1993).

\bibitem{Kyu1}T. Kyu, I. Ilies, and M. Mustafa, Journal de Physique iv, colloque. {\bf 3}, 37 (1993).

\bibitem{Taylor}J.-C. Lin and P. L. Taylor, Mol. Cryst. Liq. Cryst. {\bf 237}, 25 (1993).

\bibitem{Teixeira}P. I. C. Teixeira and B. M. Mulder, Phys. Rev. E {\bf 53}, 1805 (1996) and
references therein.

\bibitem{Lansac}Y. Lansac, F. Fried, and P. Maissa, Liq. Cryst. {\bf 18}, 829 (1995) and 
references therein.

\bibitem{Cahn1}J. W. Cahn and J. E. Hilliard, J. Chem. Phys. {\bf 28}, 258 (1958).

\bibitem{Cahn2}J. W. Cahn, J. Chem. Phys. {\bf 30}, 1121 (1959).

\bibitem{Cahn3}J. W. Cahn and J. E. Hilliard, J. Chem. Phys. {\bf 31}, 688 (1959).

\bibitem{Langer}J. S. Langer, M. Bar-on, and D. Miller, Phys. Rev. A {\bf 11}, 1417 (1975).

\bibitem{Binder}K. Binder and D. Stauffer, Phys. Rev. Lett. {\bf 33}, 1006 (1974).

\bibitem{Siggia}E. Siggia, Phys. Rev. A {\bf 20}, 595 (1979).

\bibitem{Kyu2}C. Shen and T. Kyu, J. Chem. Phys. {\bf 102}, 556 (1995).

\bibitem{Ginzburg96}V. V. Ginzburg, L. Radzihovsky, and N. A. Clark, Phys. Rev. E (submitted).

\bibitem{Coniglio}A. Coniglio and M. Zannetti, Europhys. Lett. {\bf 10}, 575 (1989).

\bibitem{Glotzer}S. G. Glotzer and A. Coniglio, Phys. Rev. E {\bf 50}, 4241 (1994).

\bibitem{Castellano}C. Castellano and M. Zannetti, Phys. Rev. E {\bf 53}, 1430 (1996).

\bibitem{Lifshitz}I. M. Lifshitz and V. V. Slyozov, J. Phys. Chem. Solids {\bf 19}, 35 (1961).
\end{thebibliography}
\end{document}